\documentclass{mn2e}

\usepackage[dvips]{graphicx}

\begin{document}

\title[Building a control sample for galaxy pairs]{Building a control sample for galaxy pairs}
\author[Perez et al.]{Josefa Perez$^{1,2}$, Patricia Tissera$^{1}$, Jeremy Blaizot$^{3}$\\
$^{1}$Instituto de Astronom\'\i a y F\'\i sica del Espacio,Conicet-UBA, CC67, Suc.28,Ciudad de  
      Buenos Aires, Argentina.\\
$^{2}$Facultad de Ciencis Astron\'omicas y Geof\'\i sicas, Universidad Nacional de
      La Plata, Argentina.\\
$^{3}$Universit\'e de Lyon, Lyon, F-69003, France ; Universit\'e Lyon 1,
      Observatoire de Lyon, 9 avenue Charles Andr\'e, Saint-Genis Laval, F-69230, France}

\maketitle

\begin{abstract}

Several observational works have attempted to isolate the
effects of galaxy interactions by comparing galaxies in  pairs with isolated galaxies. 
 However, different authors have proposed different ways to build these so-called control samples (CS). 
By using mock galaxy catalogues of the SDSS-DR4 built up from the Millennium Simulation, we explore how the way of building a CS might introduce biases
which could affect the interpretation of results.  
We make use of the fact that the physics of interactions is not
included in the semianalytic model, to infer that any difference between the mock control 
and pair samples can be ascribed to selection biases. 
Thus, we find that galaxies in  pairs artificially tend to be older and more bulge-dominated, and to have
less cold gas and different metallicities than their isolated counterparts. Also because of a biased selection, 
galaxies in pairs tend to live in higher density environments, and in  haloes of larger masses. 
We find that imposing constraints on redshift, stellar masses and local densities diminishes the
selection biases by $\approx 70\%$. 
Based on these  findings, we suggest observers how to
build an unique and unbiased CS in order to reveal the effect of galaxy interactions.   

\end{abstract}
\begin{keywords}cosmology: theory - galaxies: formation - galaxies: evolution - galaxies: interactions.
\end{keywords}

\section{Introduction}

Galaxy interactions have been found to drive strong changes in the physical properties
of galaxies.  
Their effects on galaxy properties such as star formation, morphology, metallicity 
 have been largely studied in
optical (e.g. Larson \& Tinsley 1978; Donzelli \& Pastoriza 1997; Barton, Geller \& Kenyon 2000, Kewley et al. 2006)
 and infrared observations (e.g. Sanders \& Mirabel 1996;  Lin et al. 2007; Geller et al 2006).
Numerical simulations have provided insights on the relevance of mergers and interactions in 
the formation and evolution
of galaxies (Toomre \& Toomre 1972; Barnes \& Hernquist 1992; Mihos \& Hernquist 1996), principally in 
a hierarchical clustering scenario (e.g. Tissera 2000; Somerville et al. 2001; Perez et al. 2006).

With the aim to isolate the effects of interactions, it has become popular to build CS to confront
 the properties of galaxies in pairs.
 Lin et al. (2007) found that the infrared luminosity  of blue merging galaxies and kinetically
close pairs (for a given stellar mass) almost duplicates the infrared luminosity 
of CS randomly drawn from blue isolated galaxies. 
Using spectroscopy and infrared photometry, Geller et al. (2006) 
found a strong correlation for galaxy pairs between the Balmer decrement and the H-K colour, 
which indicates that there is an intrinsic reddening associated to the
near infrared emission  of hot dust present in  
tidally triggered-star forming regions. They also show that 
the near infrared colour  diagram is
a good indicator of interaction effects, with a larger dispersion in the H-K
colours for galaxy pairs than for control galaxies. Even more, they
found that this dispersion in the NIR colour diagram for galaxy pairs increases 
for smaller relative projected separations.
In the optical,  
 De Propris et al. (2005)  showed  that interacting galaxies  in the Millennium Galaxy Catalogue
 tend to be marginally bluer than non-interacting galaxies. 
They also found that galaxy pairs have a larger contribution of very early and very late type objects 
with respect to their  control galaxies. They interpreted these facts as the result of the action of mergers and interactions on the triggering of star formation and morphology evolution.

Large galaxy surveys such as the  2dF Galaxy Redshift Survey (2dFGRS., Colless et al. 2001)
and Sloan Digital Sky Survey (SDSS, York et al. 2000) allow a statistical and comprehensive 
 study of different properties (i.e.
star formation activity, morphology) for  galaxies with and without a close companion.
Close interactions at
low relative velocity  have been found to trigger significant star formation activity 
(e.g. Lambas et al. 2003; Nikolic et al. 2004; Luo et al. 2007; Li et al. 2008). 
In fact, the mean specific star formation rate 
of galaxy pairs with projected separations lesser than $\sim 30 \,\rm kpc$ 
is significantly enhanced over the mean value corresponding to galaxies without a close companion, 
inhabiting similar  environment (Lambas et al. 2003; Alonso et al. 2004; Alonso et al. 2006).
It has been also found that galaxy interactions might induce star formation in all environments  
(Alonso et al. 2004; Li et al. 2007). 
In addition, the analysis of colours for galaxies in pairs  shows that, although close
pairs have a larger fraction of blue galaxies, they also exhibit an excess of
red galaxies with respect to those systems without a close companion located in regions of 
similar densities (Alonso et al. 2006). While the blue excess is  
associated to  systems with intense star formation triggered by the interaction, 
 the red one could be related to 
an old dominating stellar population 
or to the result of dust stirred up 
during the encounter which could hide part of the current star formation activity.

The reliability  of these outcoming results  depends on the details
of the construction of these CS used for comparison. 
Different authors resorted to different way of building up CS with the aim
at isolating the effects of interaction. 
 Barton et al (2007)  noticed that galaxies in close pairs 
reside preferentially within cluster or group-size haloes, representing
 a biased population, not suited for direct comparison to field galaxies.   
In order to isolate the effect of interactions, these authors suggest a construction of 
a clean pair sample built with galaxy pairs which are isolated 
in their dark matter haloes and,  for comparison, a control
sample populated only with one isolated galaxy  in the halo. 
Lambas et al. (2003) removed galaxies
in groups and clusters from the 2dFGRS by cross-correlating the catalogue with the groups sample of
Merchan \& Zandivarez (2002), before selecting galaxies in pairs and in the control sample.
 However, a comprehensive study of the possible  biases that 
could affect the results is still missing.

In this paper,  
we use a mock galaxy catalague of the SDSS-DR4  to carry out a global study of 
 biases which could arise in 
  the selection of  control galaxies and to suggest how to
build an unique and unbiased CS  to isolate the effect of interactions.  
The comparison with observations will be carried out in a separate paper.

The galaxy pair catalogue studied in this paper was built up from 
a mock catalogue of the  SDSS-DR4  constructed  from  the galaxy sample generated by
 the semi-analytic model (SAM) of De Lucia \& Blaizot (2007)
applied to  one of the largest N-body cosmological simulation, the so-called  Millennium  Simulation.
The SAM does not include the physics of interactions, hence, when selecting pairs and CS, any difference
in their properties cannot be attributed to interactions but to the constraints used to build the
CS. We will make profit of this fact to obtain the criteria to build up a proper CS which 
univocally allows the individualization of the effects of interactions.

 This paper is organized as follows. Section 2 describes 
the semi-analytical model used to build the mock galaxy catalogues 
from where the galaxy  pair  and control samples are selected.  
The analysis of different bias effects in the selection of
CS is shown in Section 3. 
In Section 4, we discuss how to correct these biases in order 
to build a suitable CS. We  suggest the observers
 how these findings could be taken into account in real surveys.
 An example of this procedure is shown in Section 5,
where we use the  theoretical analysis 
of the mass-metallicity relation to infer  possible biased 
results from observations. 
Conclusions are summarized in  Section 6.

\section{Mock galaxy pair catalogue.}

We use the catalogue of galaxies built up by De Lucia \& Blaizot (2007) from the 
Millennium Simulation (Springel et al. 2005).
This simulation describes the evolution of the  dark matter component assuming
 a $\Lambda$CDM cosmology with  cosmological parameters 
determined from the combined analysis of the 2dFGRS (Colless et al. 2001)
and the first year WMAP data (Spergel et al. 2003):
 $\Omega_{m}=0.25$, $\Omega_{b}=0.045$, $\Omega_{\Lambda}=0.75$, 
$H_{0}=100\,h$, $h=0.73$, $n=1$ and $\sigma_{8}=0.9$.
The Millennium Simulation follows $N= 2160^3$ particles with mass $8.6 \times 10^8 {\rm h^{-1}} M_{\odot}$ within a 
comoving periodic box of $500 {\rm h^{-1} Mpc}$ on a side. 
In a large simulation like the present one,
 a rich substructure of gravitationally bound dark matter subhaloes
is found to orbit within larger vitalized haloes. Then, the  identification of substructure
is a complex process which  required sophisticated tools specially designed to select
subhaloes within  larger haloes in an efficient way (Springel et al. 2001).
After a gravitational binding analysis,  only bound substructures with more than 20 particles
 are included as subhaloes ($1.7 \times 10^{10} {\rm h^{-1}} M_{\odot}$).

All physical processes associated to the
baryonic matter are described by phenomenological prescriptions parametrized to match
observed galaxy properties like
luminosity and colour distributions, morphologies, gas
and metal contents as explained  in detail by De Lucia \& Blaizot 
(2007; see also Croton et al. 2006).
The adopted SAM models the star formation,
generation of galactic winds, supernova feedback, 
black hole growing and also the  suppression of cooling flows
by AGN feedback. 
However, the SAM treats galaxy mergers as an instantaneous process, and does
not include pre-merger star formation induced by tidal interactions. As a consequence,
galaxies which are about to merge in the model (i.e. galaxy pairs) do not show any 
 signatures of interaction in their astrophysical or morphological properties.
Colours and magnitudes are estimated by adopting the population synthesis models
of Bruzual and Charlot (2003), and 
are dust corrected following 
Guiderdoni \& Rocca-Volmerange (1987) as explained by De Lucia \& Blaizot (2007).

Thus, the synthetic galaxy catalogue (hereafter MR galaxies) provides information on
star formation rates, total stellar masses ($M_{*}$), SDSS photometric magnitudes,
black hole mass, masses in cold and hot gas phases, masses in metals
in the different baryonic components and also dark matter halo masses.

In the SAM, galaxies are  classified as: 
central galaxies  of dark matter haloes (Type 0) or satellites  (Type 1 and 2). 
 Type 1 satellites  inhabit dark matter subhaloes within  larger ones  
while Type 2 satellites  have lost their own dark matter haloes
as they entered into  larger ones. 
After losing its subhalo, positions and velocities of Type 2 satellite galaxies,
are determined by those of the most bound particle of the subhalo
at the last time it was identified. At this point, 
the satellite galaxy merges with a central galaxy after a certain  merging time
estimated by using the dynamical friction model (Binney \& Tremaine 1987). 

Thus, combining the large dark matter simulation and the  SAM, it is possible to track the
evolution of galaxies throughout volumes comparable to 
the largest current galaxies surveys such as the SDSS.

\subsection{Mock galaxy pair and the basic control sample.}\label{samples}

A reliable confrontation between  observations and models requires a correct 
mimic of the observational procedure. We use MoMaF (Blaizot et al. 2005) 
to create a SDSS-DR4 mock catalogues from MR galaxies. These
mocks allow us to select simulated galaxy with the same
set of {\it observational} criteria as in Alonso et al. (2006):
$0.01 < z < 0.1$ and $r < 17.77$. 
From this  redshift and r-magnitude limited sample made of 254335 galaxies, 
we search for galaxy pairs imposing thresholds in projected separation   
 ($r_{\rm p}<100 \rm kpc$) and relative radial velocity ($\Delta cz <350$ km s$^{-1}$)
 (Lambas et al 2003; Alonso et al 2004; Alonso et al 2006). We obtained
 a  Pair Catalogue composed by 37590 galaxies. 
The remaining  galaxies without a close companion 
within the adopted thresholds will constitute the Non-Pair Sample (NPS).

We calculate the local environment of galaxies  by estimating the 
 local projected density parameter defined
as $\Sigma=5/(\pi d^{2})$, where  $d$ is  the projected distance
to the 5$^{th}$ nearest neighbour brighter than M$_{\rm r} = -20.5 $, with
$\Delta cz < 1000$ km ${\rm s^{-1}}$ (Balogh et al.2004; Alonso et al.2006). 
The limits on redshift and r-band magnitude have been imposed over the pair
and control galaxies, so that both are equally affected by incompleteness 
problems.

 As a first order CS, we select galaxies in the NPS by requiring them to  have the  same
redshift  and absolute r-magnitude distributions than those in the Pair Catalogue. Thus,
 for each galaxy in a pair, 
we look for a NPS galaxy with  the same redshift and r-magnitude 
but without a near companion in order to build up  the first CS (hereafter, {\bf Control 1}). 
In Table~\ref{tablacontrol} we summarize the constraints applied to build up all the control samples discussed below.

\begin{table}
\caption{Control Samples: constraints applied to build up the analysed control samples.
}\label{tablacontrol}
\centering
\begin{tabular}{cccccccc} 
\hline \hline
Control & $L$ & $z$ & $M_*$ & $\Sigma$ & $B/T$ &  $M_{\rm halo}$ & Galaxy Type\\
\hline
Control 1 & X &  X  &     &          &           &    &    \\

Control 2 &  &  X     &  X   &          &           &  &    \\

Control 3 &  &  X     &  X   &   X       &           &  &    \\

Control 4 &  &  X     &  X   &   X       &  X         &  &    \\

Control 5 &  &  X     &  X   &   X       &  X         &  X&    \\

Control 6 &  &  X     &  X   &   X       &  X         &  &X    \\

Control 7 &  &  X     &  X   &          &          &X  &    \\
\hline
\end{tabular}
\end{table}


\section{Analysis of possible bias effects in the control sample.}

Taking profit of the fact that the model does not include the physics of interactions,
we expect  galaxies in pairs and in the CS to have the same properties, at least, if we
suppose that they have experienced the same average history of assemble. 
So, any differences should be ascribed to bias effects in the selection of the pair sample, not to interactions.
We check this hypothesis particularly focusing on the analysis of
colour and cold gas distributions for galaxies in pairs and in the CS. We use
these relations because they should be strongly connected with any possible 
star formation activity triggered by  interactions.

\begin{figure*}
\centering
\includegraphics[width=6.5cm,height=5.5cm]{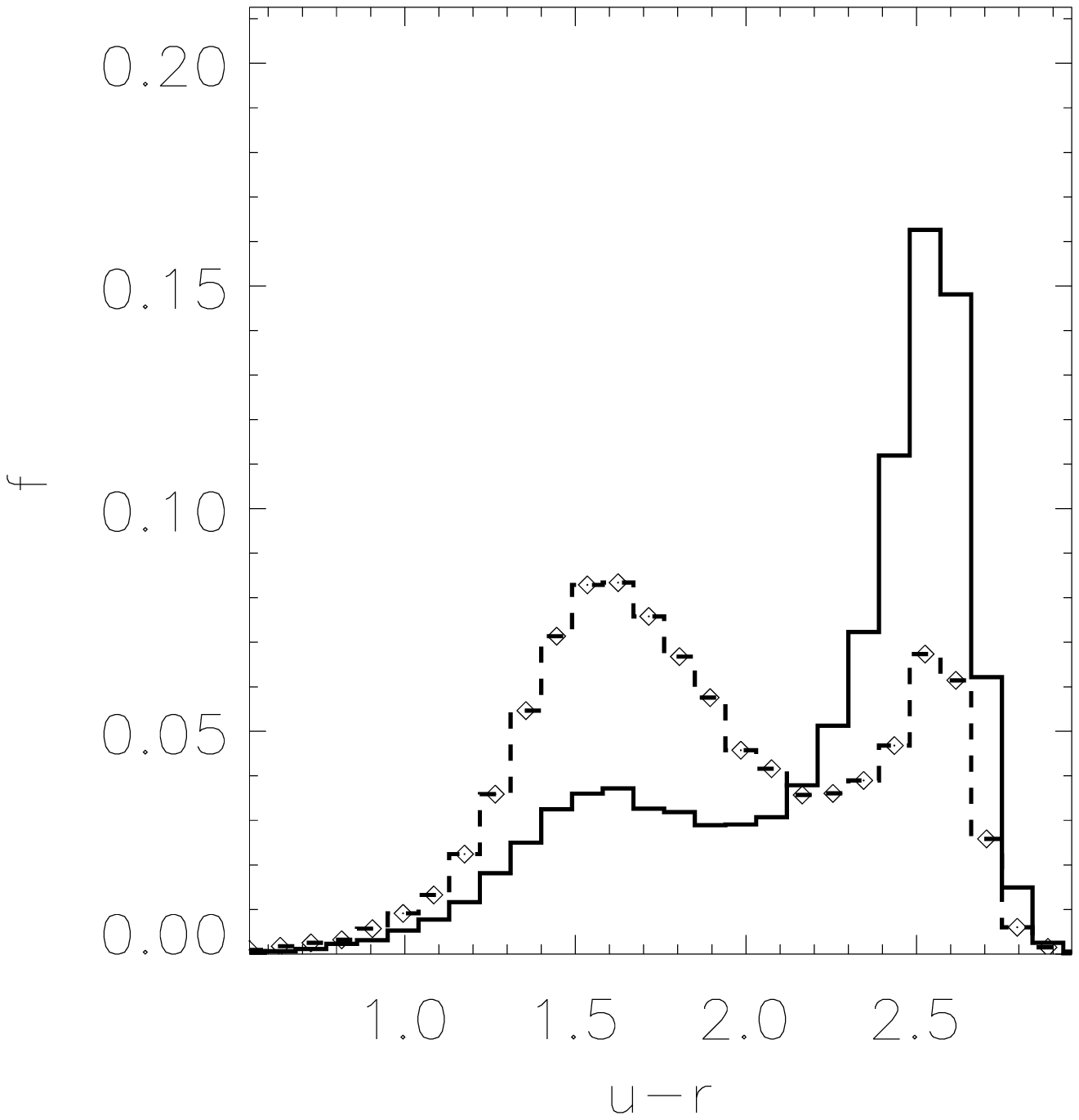}
\includegraphics[width=6.5cm,height=5.5cm]{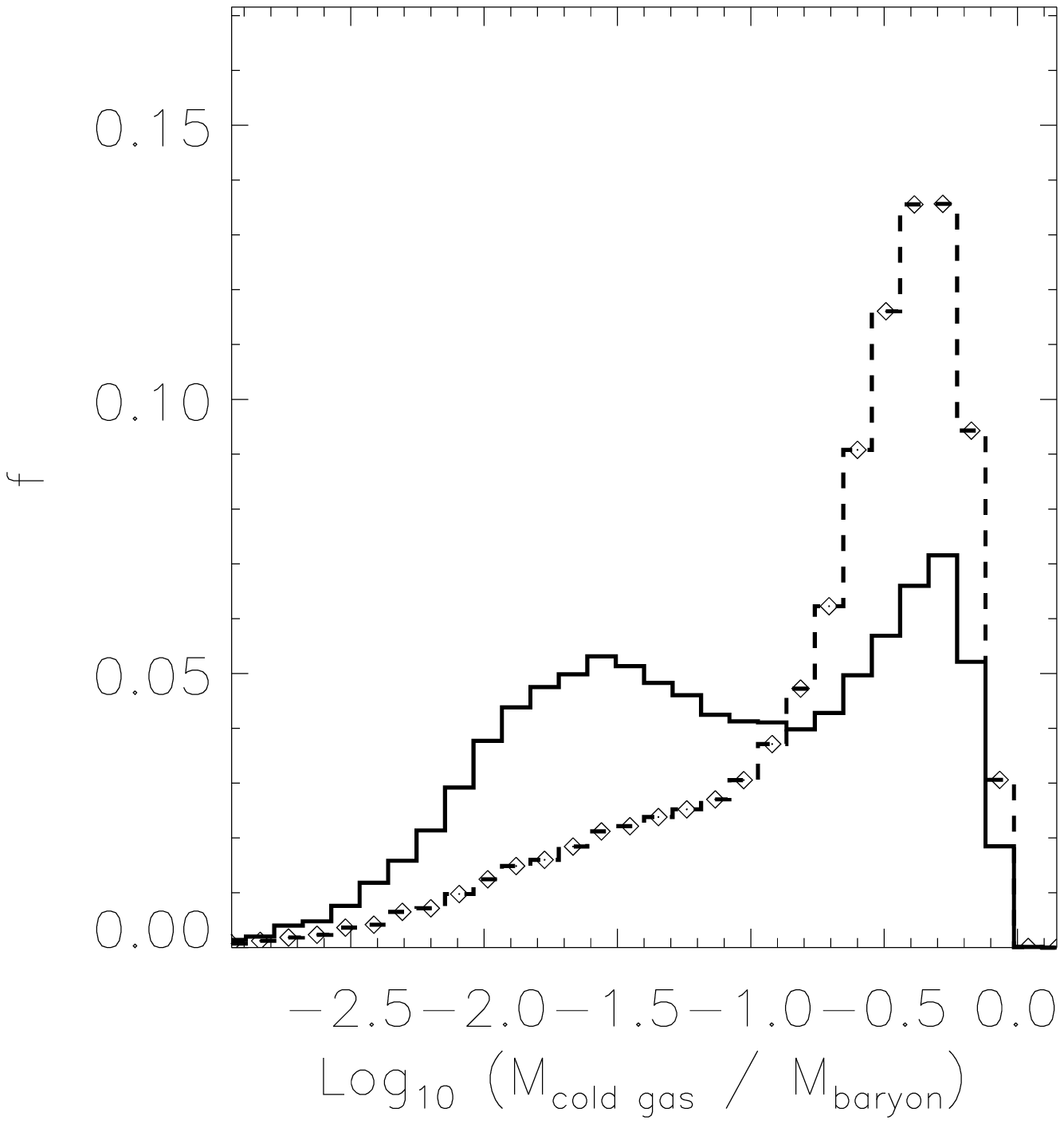}
\caption{
The (u-r) colour distributions (left panel) and cold gas fraction distributions (right panel) 
for both galaxies in pairs (solid line) and
 in the {\bf Control 1} (dashed line). Error bars are standard deviations computed for 100 realizations
of  control samples (see the text for more details).}
\label{ur1}
\end{figure*}

\begin{figure*}
\centering
\includegraphics[width=6.5cm,height=5.cm]{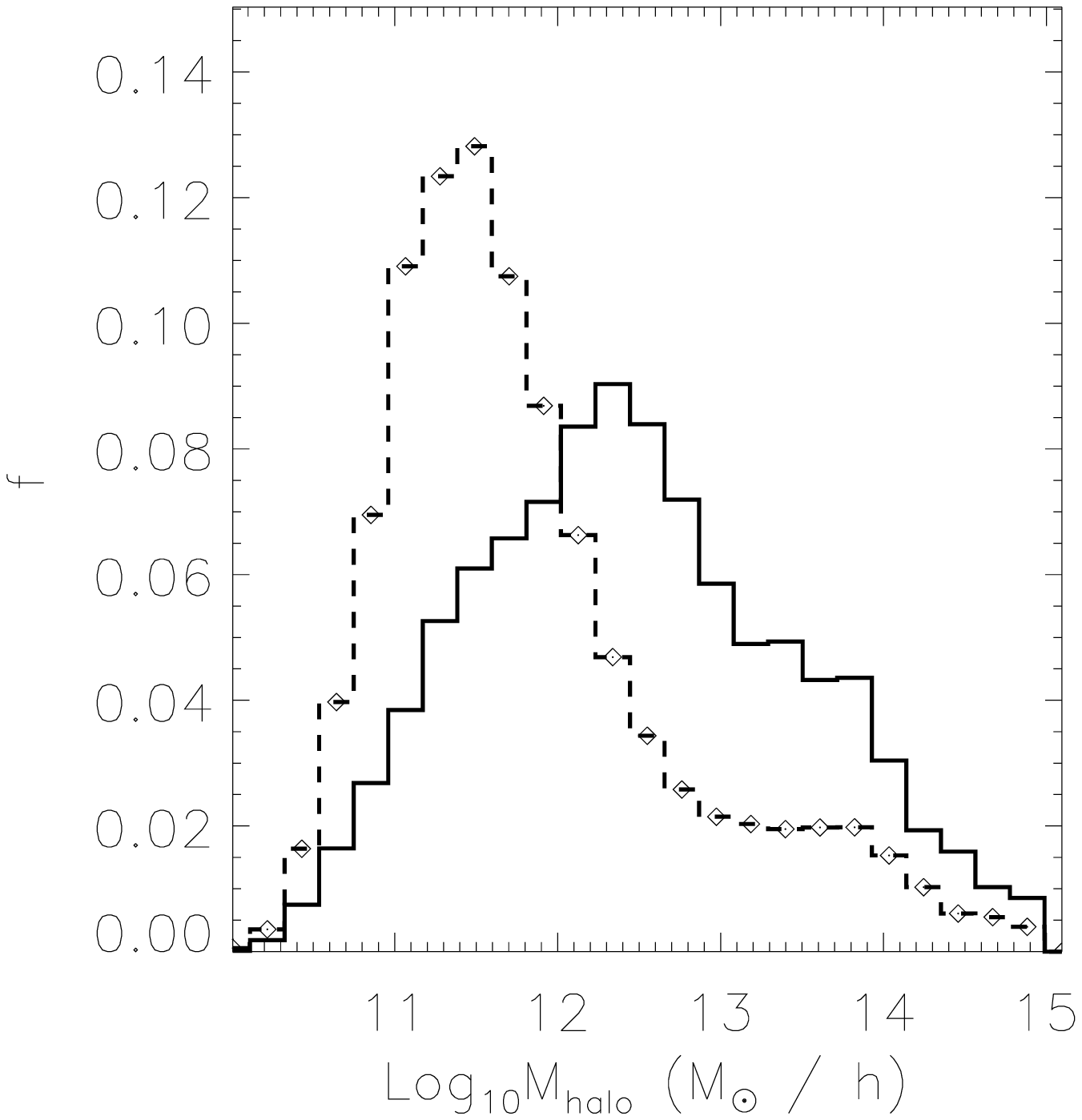}\includegraphics[width=6.5cm,height=5.cm]{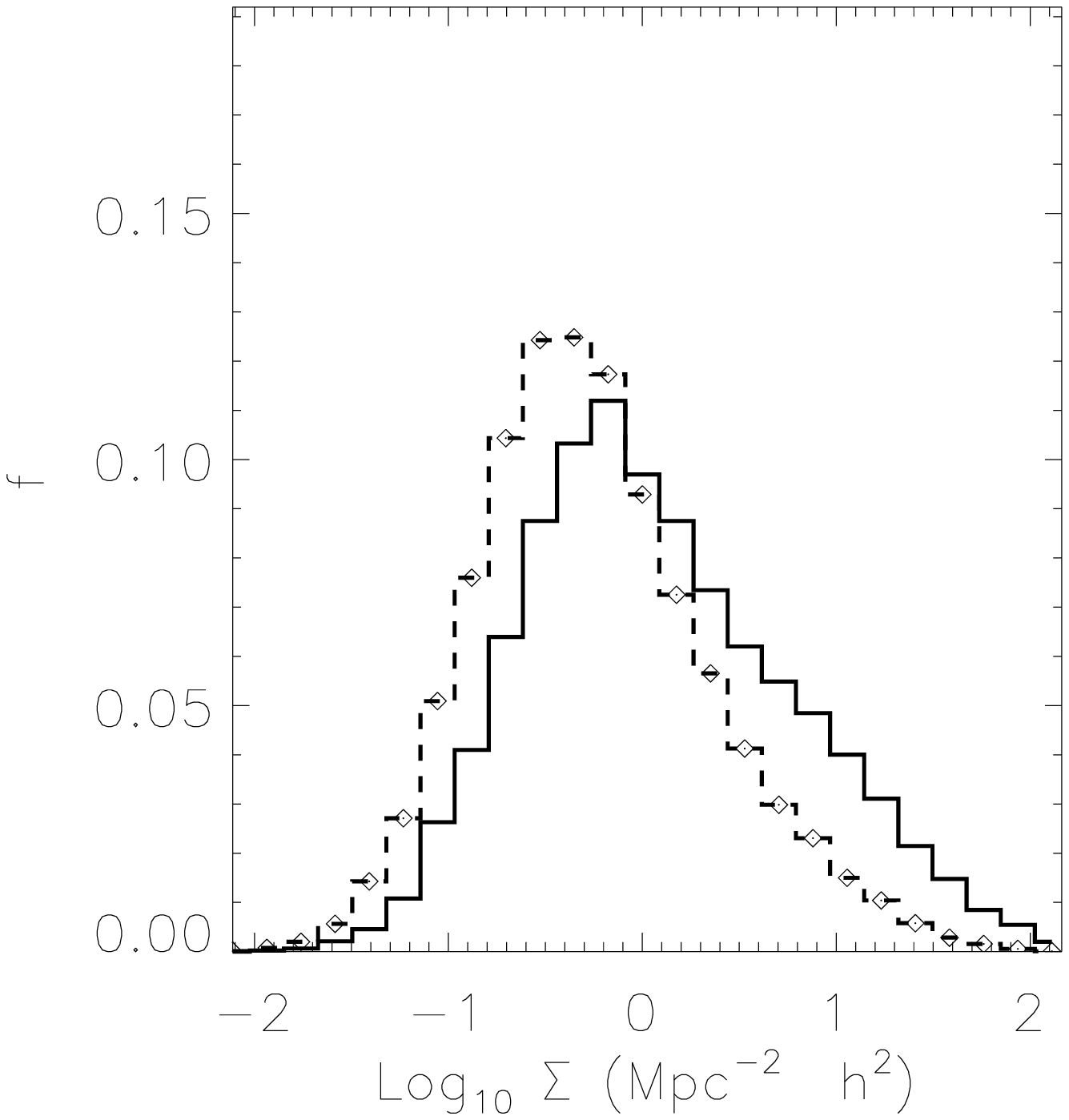}\\
\includegraphics[width=6.5cm,height=5.cm]{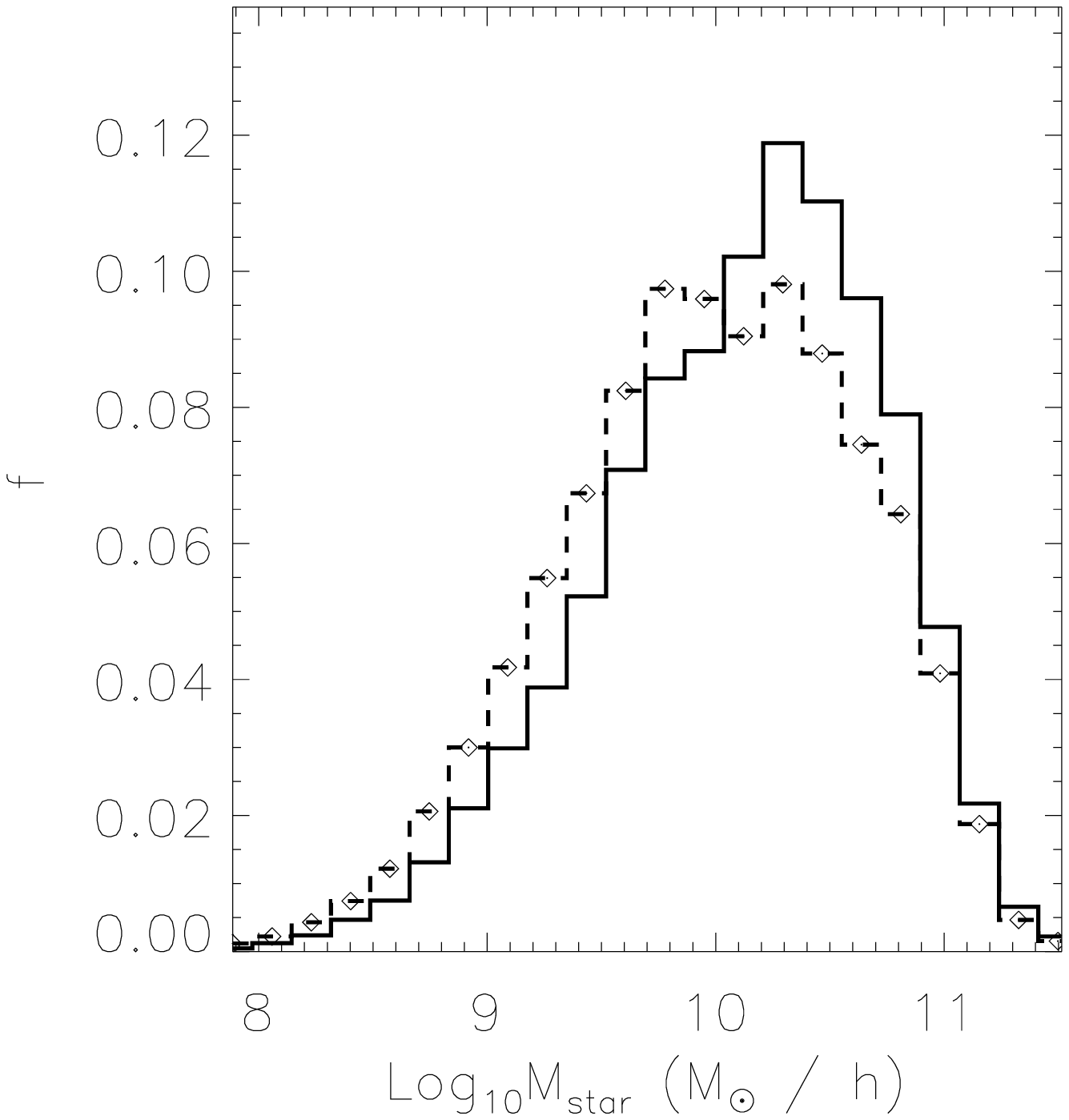}\includegraphics[width=6.5cm,height=5.cm]{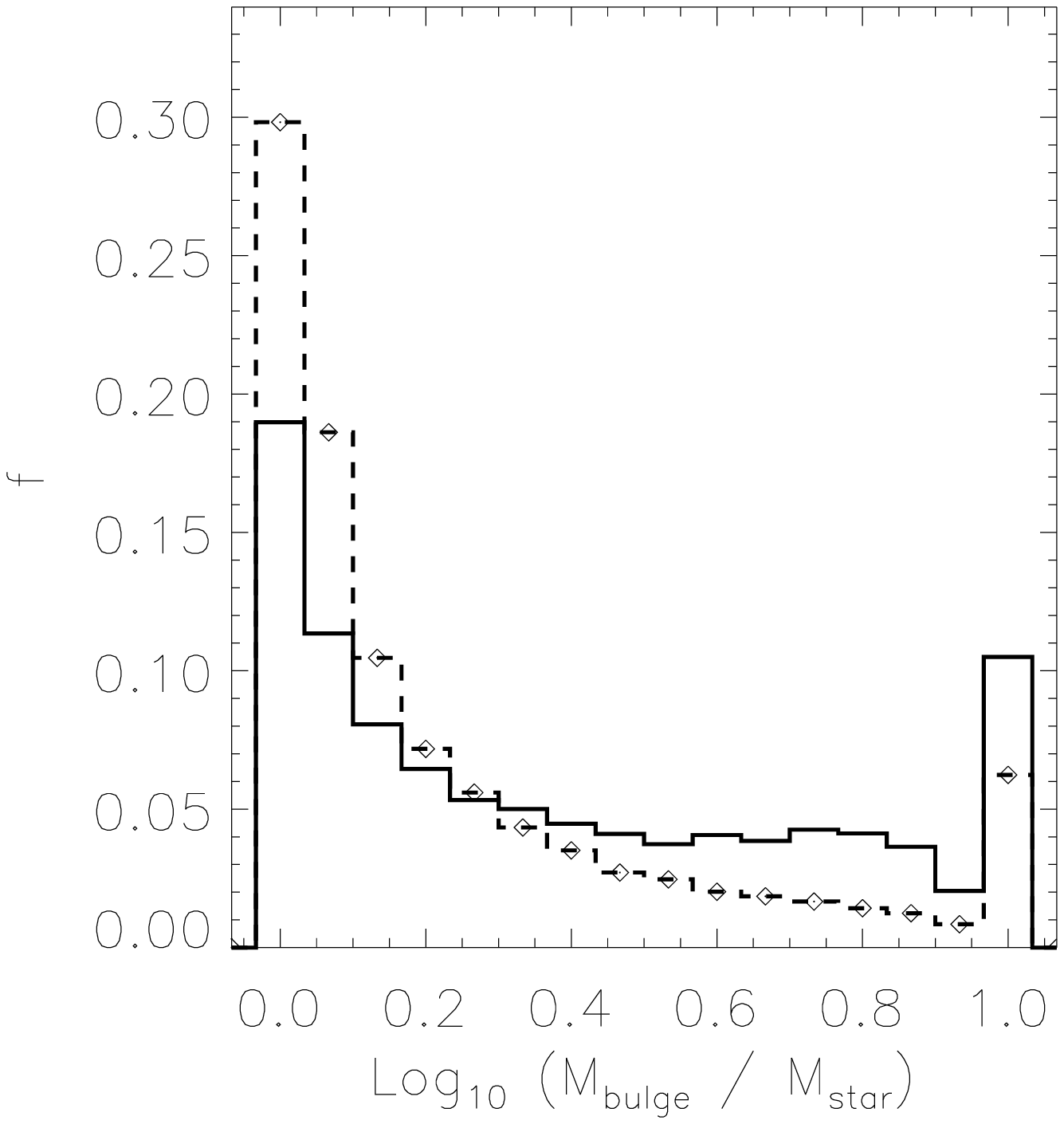}
\caption{ Histograms of the  dark matter halo masses (upper left panel),
local density parameter ($\Sigma$) distributions (upper right panel), 
stellar mass distributions  (lower left panel) and, 
morphological distributions, represented by the bulge to total stellar masses ($B/T$) parameter (lower right panel), 
for  galaxies in pairs (solid line) and in the {\bf Control 1} (dashed line).}
\label{bias1}
\end{figure*}

Comparing the colour and cold gas fraction distributions for galaxies in pairs
and in the {\bf Control 1} (Fig.~\ref{ur1}), we can appreciate significant differences between both samples 
which can not be attributed to the effect of interactions as explained before. 
Pairs exhibit an important excess of red and a deficit of blue galaxies 
(and consistently, a lower cold gas fraction) compared to  the {\bf Control 1}. 
Other physical properties of galaxies with and without a close companion are compared, such us: 
 halo masses ($M_{halo}$), local density environment ($\Sigma$), stellar masses ($M_{*}$) 
and bulge-to-total ($B/T$) ratio (Fig.~\ref{bias1}).  
We find that the dark matter halo distribution (the most difficult property to measure observationally) 
 is the one that exhibits the largest bias. In agreement with Barton et al. (2007), 
we find that galaxies in pairs tend to belong to larger haloes than galaxies in the {\bf Control 1}. 
A less observationally demanding  way to assess  the role of  
environments in driving bias effects is by using the local projected density estimator,
 $\Sigma$. In agreement with previous results (e.g. Lambas et al. 2003; Alonso et al. 2004),  
 we find that galaxy pairs tend to inhabit higher density regions that their isolated counterparts in the {\bf Control 1}.  
Beside these environmental biases, the figure also shows that galaxies in   pairs   tend slightly to
have larger stellar masses  and more important bulges (i.e. larger  bulge to total stellar mass, $B/T$)
 than galaxies in {\bf Control 1}. Although, these effects are less important, we have to take them  into account 
in order to select a suitable CS. We note however  that, in hierarchical clustering scenarios,  
 larger stellar masses systems have larger probability to have
grown by mergers, which in the SAMs directly  fed the bulges. So, in our samples these two parameters 
 are very closely related.

\begin{figure}
\resizebox{6cm}{!}{\includegraphics{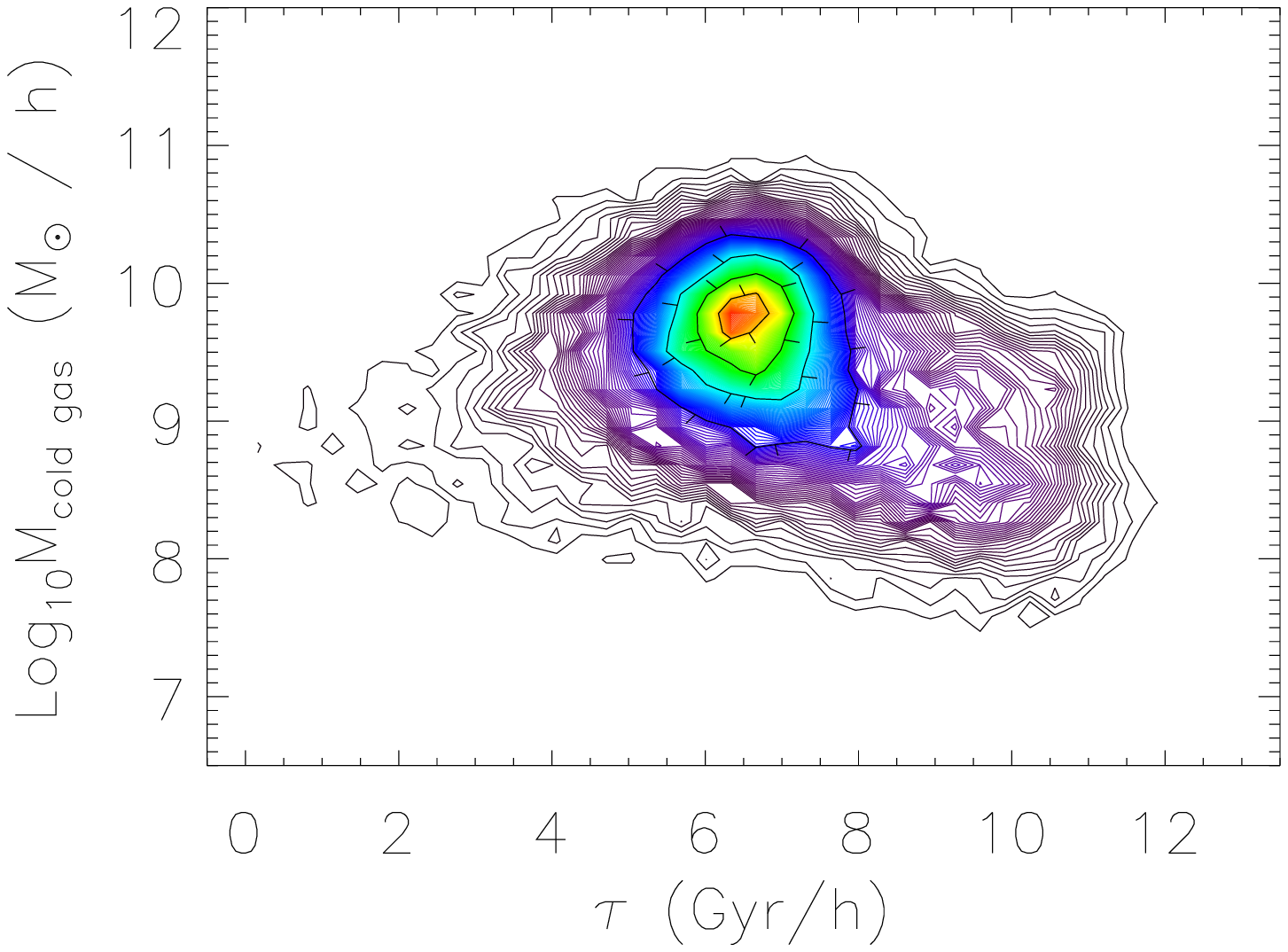}}
\hspace{+0.2cm}
\resizebox{6cm}{!}{\includegraphics{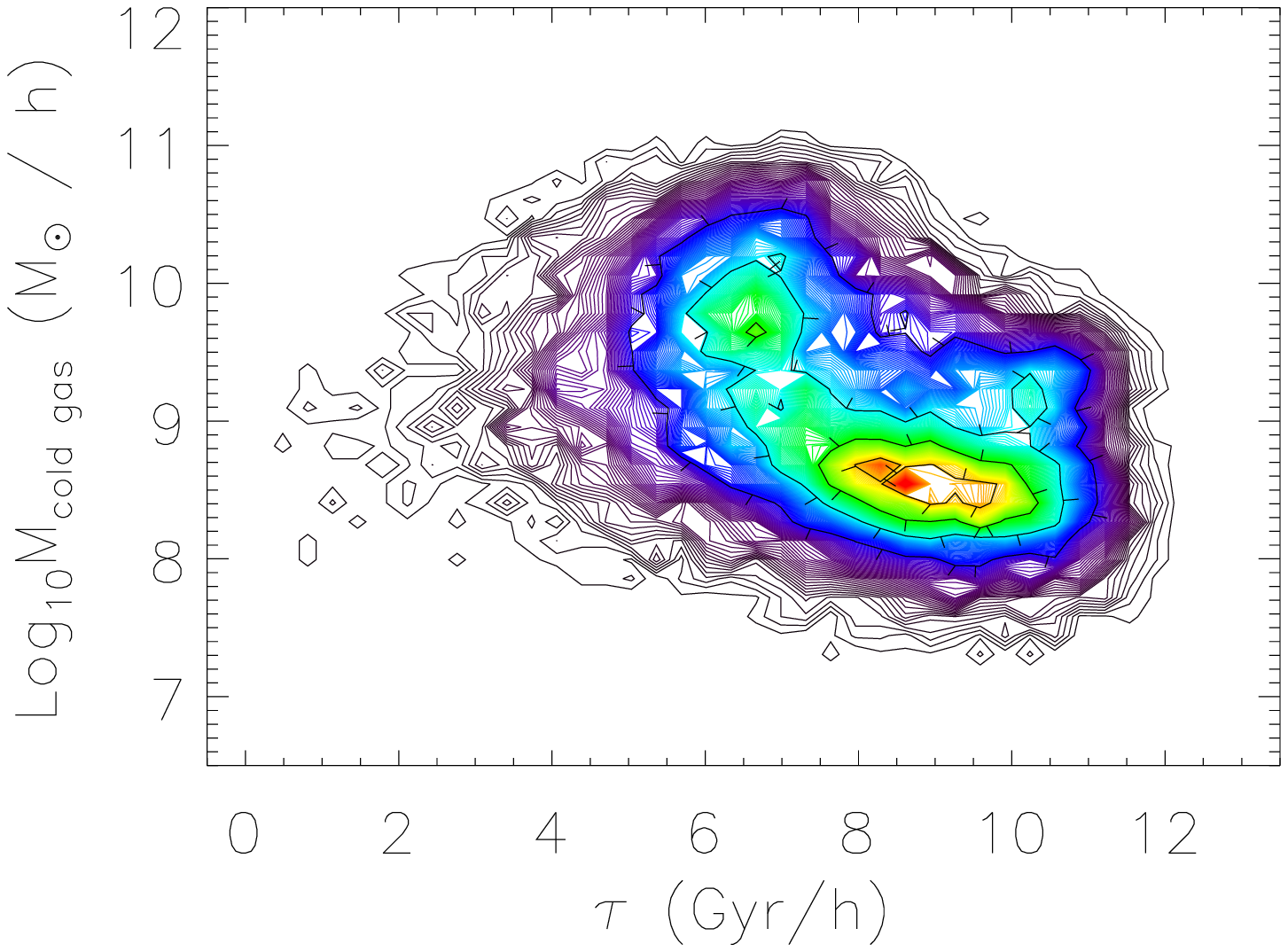}}
\hspace{+0.2cm}
\caption{
Contour plots of cold gas fraction and  age weighted stellar mass parameter, $\tau$,
for galaxies in the {\bf Control 1} (upper panel) and in pairs (lower panel). 
The sequence from red to blue colours indicates a decrease in the galaxy number density.}
\label{ur-age1}
\end{figure}

In Fig.~\ref{ur-age1}, 
we show number density of  galaxies in pairs and in the {\bf Control 1} on a 
cold gas fraction and  stellar- mass weighted age ($\tau$) plane.
Galaxies in pairs tend to be $\approx 10 \%$ older than those in the {\bf Control 1}, with
a mean value of $\tau$  equal to $8 \, \rm Gyr \, h^{-1}$ for galaxies in pairs 
and $7 \, \rm Gyr \, h^{-1}$ for galaxies in the {\bf Control 1}. 
Consistently, galaxies in pairs have less cold gas content than galaxies in the  {\bf Control 1}.
From this figure, we can also see  that galaxies in pairs have clearer bimodal distributions.

\begin{figure}
\includegraphics[width=8.2cm,height=8.6cm]{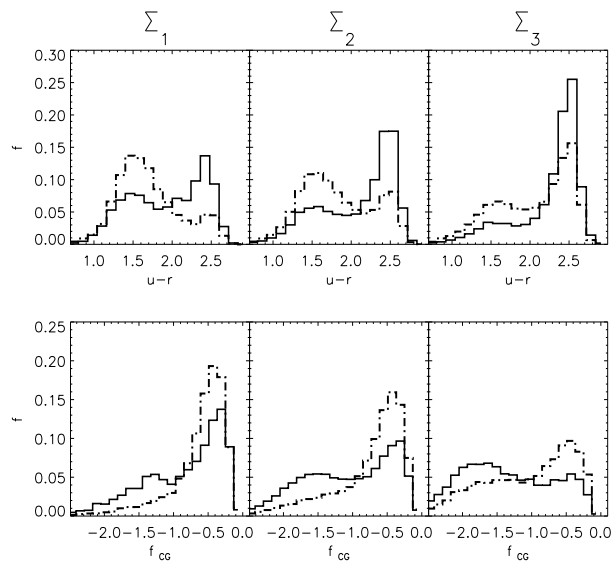}
\caption{
Histograms of colours (upper panels) and  cold gas fractions, $\rm f_{CG}$, (lower panels),
for galaxies in the {\bf Control 1} (dashed lines) and in pairs (solid lines). 
Plots have been divided in three panels corresponding to different density environments: 
$ -2.300 < \rm log \,\, \Sigma_{1} < -0.285$; $ -0.285 < \rm log \,\, \Sigma_{2} <  0.145$
and $ 0.145 < \rm log \,\, \Sigma_{3} < 3$. }
\label{cg-sigma}
\end{figure}

A more detailed inspection of  the colour and the cold gas fraction distributions for galaxies in  pairs 
and in the {\bf Control 1} as a function of the local density environment, reveals  that
the most significant difference is observed
 in low densities:  $ -2.300 < \rm log \,\, \Sigma < -0.285$ (Balogh et al. 2004). 
In such region, galaxies in  pairs exhibit
the largest excess of red and a deficit of blue galaxies with respect  
to those found in the  {\bf Control 1} (Fig.~\ref{cg-sigma}). 
Consistently, we find that in this low density,
galaxies in the {\bf Control 1} have a larger fraction  of cold gas, 
available 
for the  star formation activity responsible for their bluer colours. 
If the physics of baryons during interactions is not properly described in
the SAMs, and consequently, galaxy properties only change as a result of a merger,
why does the SAM   predict an excess of cold gas and bluer colours 
for the {\bf Control 1} respect to galaxies in pairs, particularly in low density regions.

Before analysing the other possible biases,  let us take into account the  different composition in galaxy types 
 for  both samples. 
We find that the Pair Catalogue is composed by a larger fraction of satellite galaxies
($28\%$ type 2 and  $30\%$ type 1) 
than the  {\bf Control 1} (see Table 1) which is dominated by  central types ($\approx 80\%$).
Central galaxies
can continuously replenish their cold gas reservoir available  
for star formation by cooling their hot gas component.
Type 2 satellites are galaxies which have 
lost their dark matter haloes 
and their hot gas components, so 
 they do not have a source of  gas accretion.  
Even more, Type 2 systems might have  
left-over cold gas but not enough to satisfy the 
threshold surface density to form stars adopted
in this SAM  (e.g. Kauffmann 1996; Croton et al. 2006), and in
consequence, they become passively redder and older.

The different recipes used to model both types of galaxies in the SAM are physically motivated and had been developed
to mimic the effects of global environment such as strangulation.
For the analysis in this current paper, it is important to bare this in mind and latter on, we will
assess its effect on the results.

Concluding, we find that {\bf Control 1}, 
selected at imposing only redshift and absolute r-magnitude constraints, has
younger and bluer, more cold gas enriched and more active star-forming systems 
than galaxies in the Pair Catalogue, biasing any direct comparison between them. 
The fact that, galaxies in the {\bf Control 1} tend to inhabit lower density regions
and  smaller dark matter haloes contributes  partially to this bias. We also find different compositions in stellar masses, types
and morphologies (Fig.~\ref{bias1}) which will be considered in the following sections.




\section{Isolating the interaction effects.}

In this Section, we systematically imposed constraints on stellar masses, local environments, morphologies
and halo masses to select different control samples and compared them with the Pair Catalogue to
assess the existence and importance of biases.
We also established an upper limit to  the importance of the galaxy type bias.

The constraints discussed in this paper can be also imposed on observed samples selected from
large surveys such as 2dFGRS or SDSS where the photometry and spectroscopic of galaxies are available.
However, since some of these constraints can be more difficult to impose than others, 
we introduced them progressively in order
to individualize the effects produced by each one.

\subsection{An observer's guide to unbias a control sample}

\begin{figure*}
\centering
\resizebox{6cm}{!}{\includegraphics{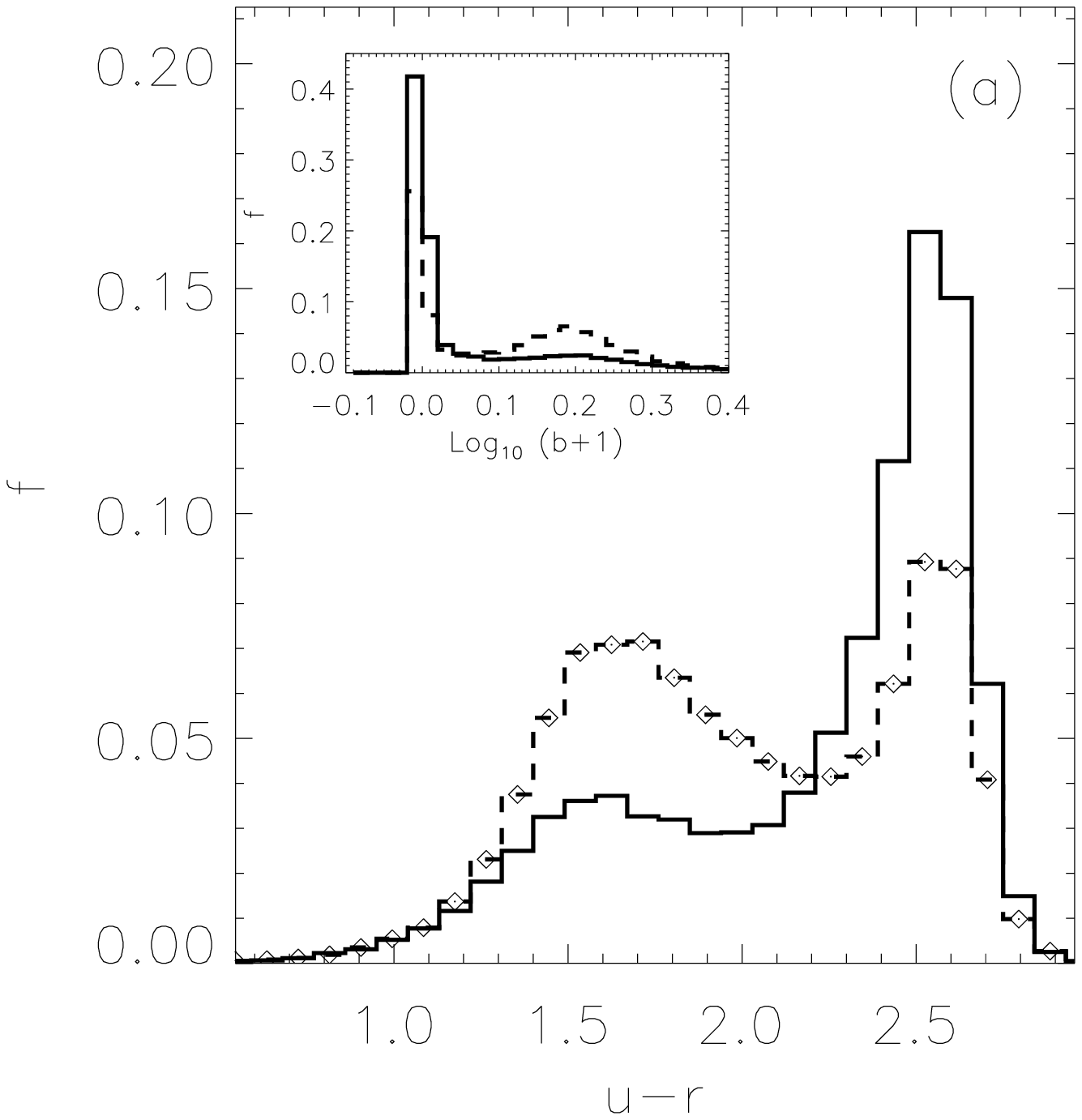}}\resizebox{6cm}{!}{\includegraphics{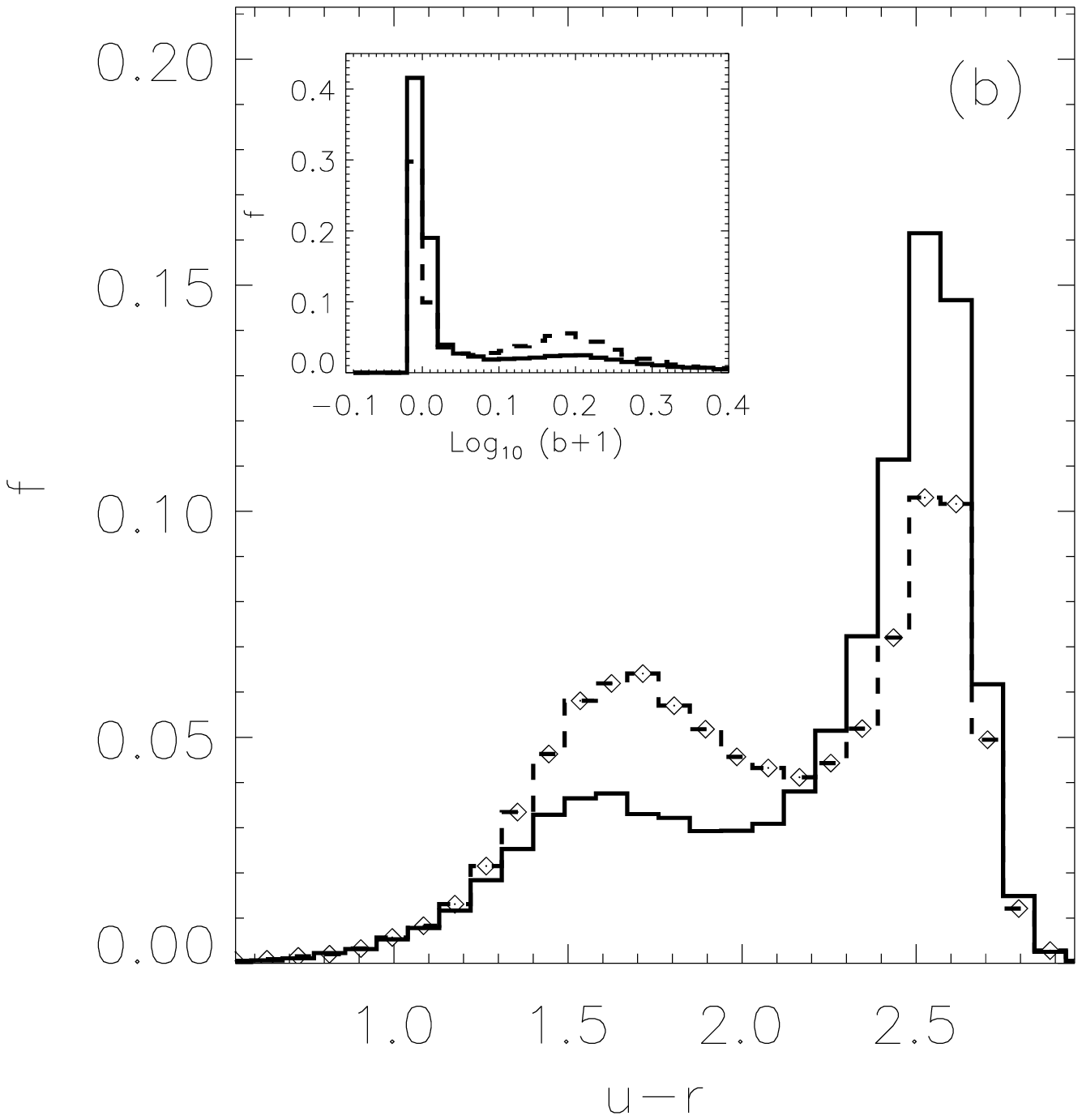}}
\hspace{+0.2cm}
\resizebox{6cm}{!}{\includegraphics{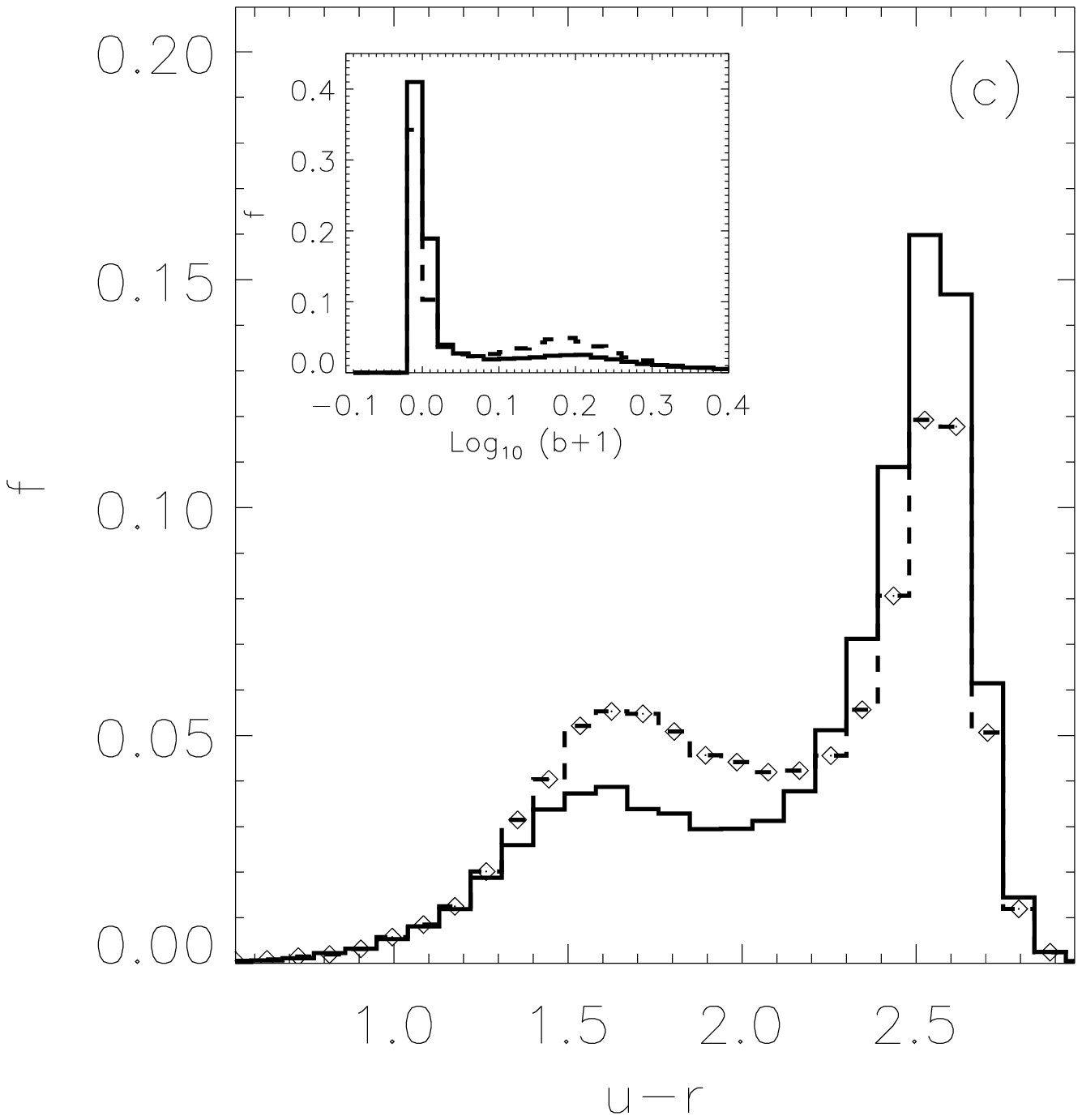}}\resizebox{6cm}{!}{\includegraphics{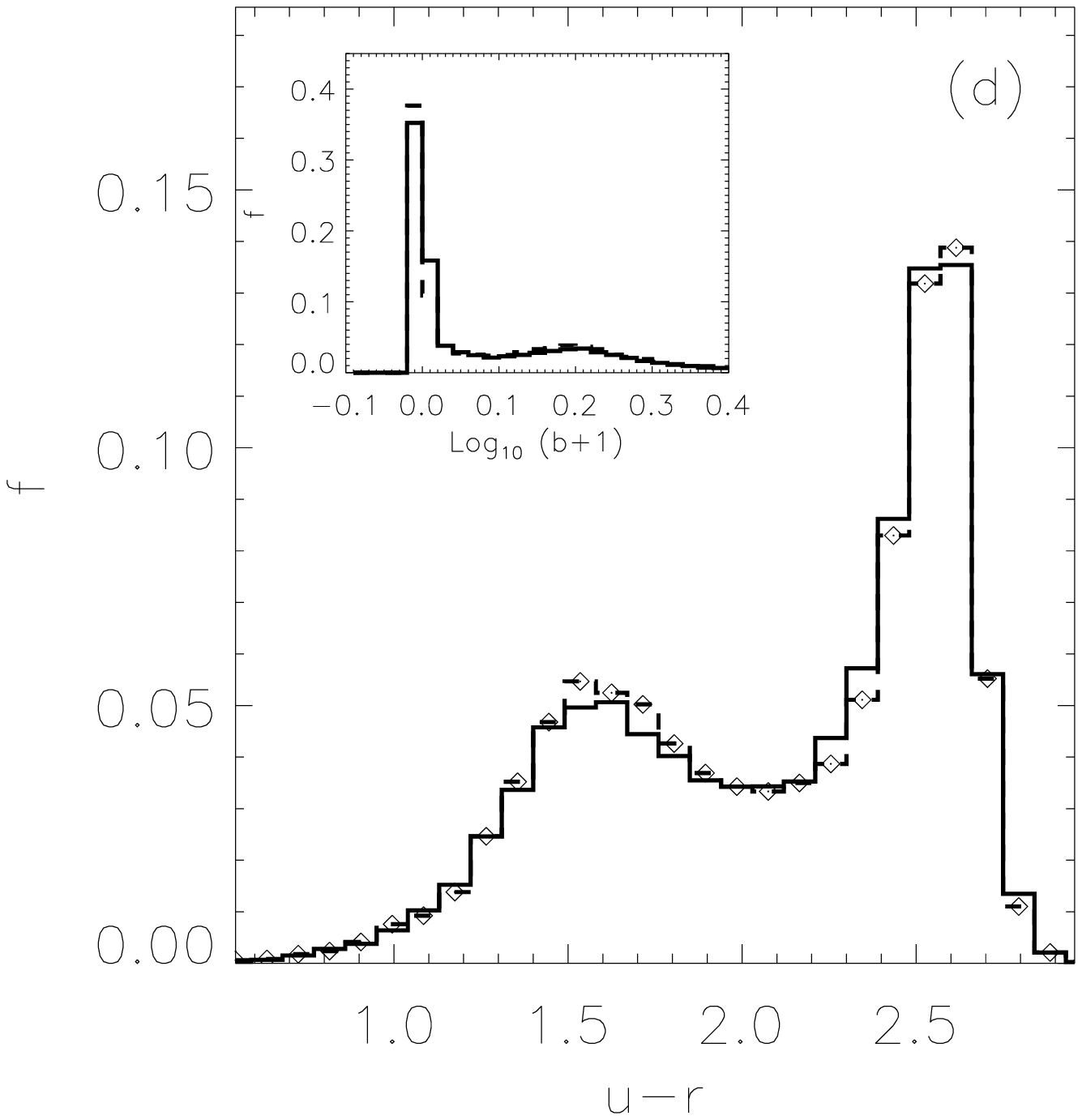}}
\hspace{+0.2cm}
\caption{
The u-r colour distributions for both galaxies in pairs (solid line) and
 without a close companion (dashed line) in {\bf Control 2} (a), {\bf Control 3} (b), 
{\bf Control 4} (c) and in {\bf Control 5} (d). 
The insets show the corresponding stellar birthrate distributions, $b$, 
for both galaxies in pairs and in control samples using the same convention of lines. }
\label{ur2}
\end{figure*}

It is widely accepted that  stellar mass is  a more fundamental quantity than luminosity 
(Kauffmann et al. 2003; Brinchmann et al. 2000; 
Baldry et al. 2006; Panter et al. 2004; Ellison et al. 2008). So, we  define 
{\bf Control 2} by selecting  NPS galaxies  which match one-to-one 
the redshift and  stellar mass distributions of galaxies in the Pair Catalogue, 
(see Ellison et al. 2008, for an exhaustive discussion). 
As it can be seen in  Fig.~\ref{ur2}a, 
the colour distributions of galaxies in {\bf Control 2}
has changed favorably in comparison to that of Control 1, diminishing 
the differences  with the colour distribution of galaxies in pairs.

We take into account the fact that galaxies in Control 1 also tend to 
be located in  lower density regions  than galaxy pairs. 
Hence, we define an alternative {\bf Control 3} 
by selecting galaxies from the NPS with redshift, stellar mass and
local density  distributions matching those of galaxies in  pairs.
 In this process,  approximately $2\%$ of the pairs samples 
can not be matched in the NPS, due to the under-representation of high
masses and high density environments in the latter. 
The colour distribution of {\bf Control 3}  (Fig.~\ref{ur2}b) shows a slight decrease and increase
of the blue and the  red peaks, respectively, with respect to the distribution of  Control 2.
Although when the agreement between {\bf Control 3} and the Pair Catalogue is better,
 discrepancies  are still present indicating that additional parameters, such us
morphology, can be considered.

In the process of improving the definition of CS, we build  the {\bf Control 4} forcing NPS galaxies 
to have an additional constraint, i.e. the morphological index ($B/T$). 
In order to define this {\bf Control 4}, almost a $6\%$ of galaxies  have been removed 
from the original Pair Catalogue, because they do not have a NPS counterparts 
which satisfies all these constraints.
As shown in  Fig.~\ref{ur2}c this new CS matches better 
the galaxy pair colours than the previous ones.

Recently, several observational methods for estimating DM halo masses have been reported. 
Spitler et al. (2008) present a method to directly estimate the total mass of a 
 dark halo  using its system of globular clusters. They show that the link between 
globular cluster systems and halo masses is independent of a galaxy type and environment, 
in contrast to the relationship between galaxy halo and stellar masses. Alternatively,  
a group finder algorithm and a dynamical mass estimation could also be used as an
observationally technique to determine halo masses. In particular, Zapata el al. (2009) 
use this technique to compare properties of galaxy groups in the SDSS-DR4 to those 
in mock catalogues. 
In consequence, it might be possible to build an observational
 CS imposing that their galaxies have the same dark matter haloes
than galaxy pairs.  Thus, in order to probe how further 
it is possible to improve the CS definition, we build the  {\bf Control 5} from NPS galaxies  
by imposing constraints on  redshift, stellar mass, 
projected local density, morphology and dark matter haloes. We note that
to build this  CS, a considerable fraction of galaxies from the original Pair Catalogue
has to be removed (approximately $40\%$) because of the lack of NPS galaxies inhabiting 
similar dark matter haloes. 
Fig.~\ref{ur2}d  shows that galaxies in the {\bf Control 5} 
fits much more closely the colour distributions of galaxy pairs than those of previous CS. 
Comparing this results with the obtained by using the original
{\bf Control 1} (Fig.~\ref{ur1}), we conclude that 
we find a suitable CS, feasibly defined in observational surveys.

Insets in Fig.~\ref{ur2} also compare the star formation activity for galaxies with and without
a near companion for each CS definition. We estimated the star formation activity 
by defining the stellar birthrate parameter,
$b=0.5 t_{H}(SFR/M_{*})$, computed as an estimator of the present star formation rate
normalized to the total stellar mass $SFR/M_{*}$ (Brinchmann et al. 2004). 
As expected, the {\it b} distributions behave consistently with those of colours.

\subsection{Additional insights from the theoretical perspective}

In the previous subsection, we discuss how to build a CS applicable to 
real galaxy surveys, and hence, potentially used by observers. Now,  we use 
parameters available only in simulations to go one step further  
analysing what we can learn from models. 

An issue to be addressed is concerning the different fractions of central and
satellite galaxies in each sample. As we mentioned before, in Control 1 there was an excess of 
central galaxy types with respect to the Pair catalogue. 
The analysis of    galaxy type populations in {\bf Control 5} shows
a significant reduction of  the fraction of central galaxies with respect to
 {\bf Control 1}: from $79.8\%$ to $57.7\%$. 
On the other hand, we have removed almost $14\%$
 of the satellite population  from the original Pair Catalogue. 
This fact implies that by  taking into account the dark matter haloes 
inhabited by galaxies ({\bf Control 5}), 
we have also removed the bias in galaxy types. Hence both the pair sample and 
the {\bf Control 5} have a final composition 
of $\sim 45\%$ of satellites and   $\sim 55\%$ of central galaxies. 
Although, this final selection on halo mass is similar in spirit to the method proposed by 
Barton et al. (2007), our criteria is less restricted because it only requires 
galaxies in pairs and in the control sample to inhabit similar mass haloes.
 
It turns out that most of the effect of the halo selection thus comes from getting
similar proportion of central and satellite galaxies in the pair and control samples. 
We have checked that indeed, replacing the halo mass condition by a condition on
galaxy type (central or satellite) yields similar results as Fig.~\ref{ur2}d. We wish to note
at this point that the semi-analytic model we are using tends to have too
steep a behavior, in the sense that satellite galaxies redden too fast  after
they enter a larger halo (Wang et al. 2007). This enhances the difference between the pair and control samples
unless they are built in a way which yields similar numbers of satellite and central galaxies.
Hence the dis-agreement found in Fig.~\ref{ur2} are somewhat over-estimated. Nonetheless, this
enhancement points us to a radical solution, also adopted to some extent by Barton et al. (2007),
which is to match halo masses and thus remove the satellite/central issue.

\begin{figure}
\resizebox{8cm}{!}{\includegraphics{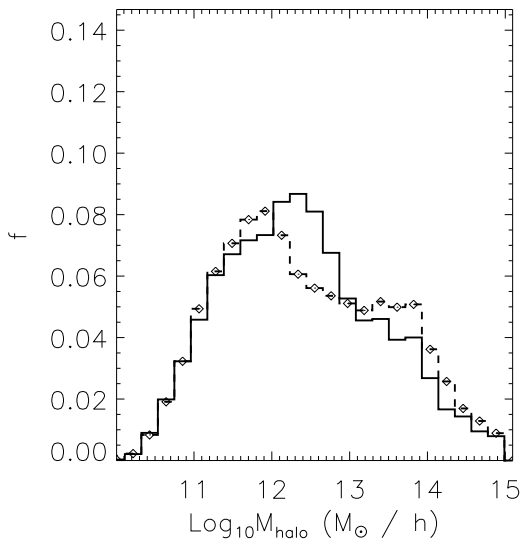}}
\hspace*{-0.2cm}
\caption{Dark matter halo distributions for galaxies in the Pair Catalogue (solid lines) 
and in {\bf Control 6} (dashed lines).}
\label{dm}
\end{figure}

In order to asses the effect of galaxy type modelling, we define {\bf Control 6 }  
by selecting galaxies from the NPS with  
similar redshift, stellar mass, local density environment, morphology type and galaxy type distributions 
to those of galaxies in pairs.
We found that in the {\bf Control 6}, $\approx 49\% $ and $\approx 52\%$ of the members are  satellite and central galaxies, respectively. These type population frequency is very similar to that found in  {\bf Control 5} where the contrain on the dark matter halo had been 
 imposed. However, the distribution of dark matter haloes of galaxies in pairs and in 
 {\bf Control 6} are still different (Fig. ~\ref{dm}). 
We claim  that the dark matter bias is a real effect 
although could be exacerbated in the SAMs so that
our results should  be considered upper limits 
(similar caution should be taken when using other models to populate haloes).

\begin{figure}
\resizebox{8cm}{!}{\includegraphics{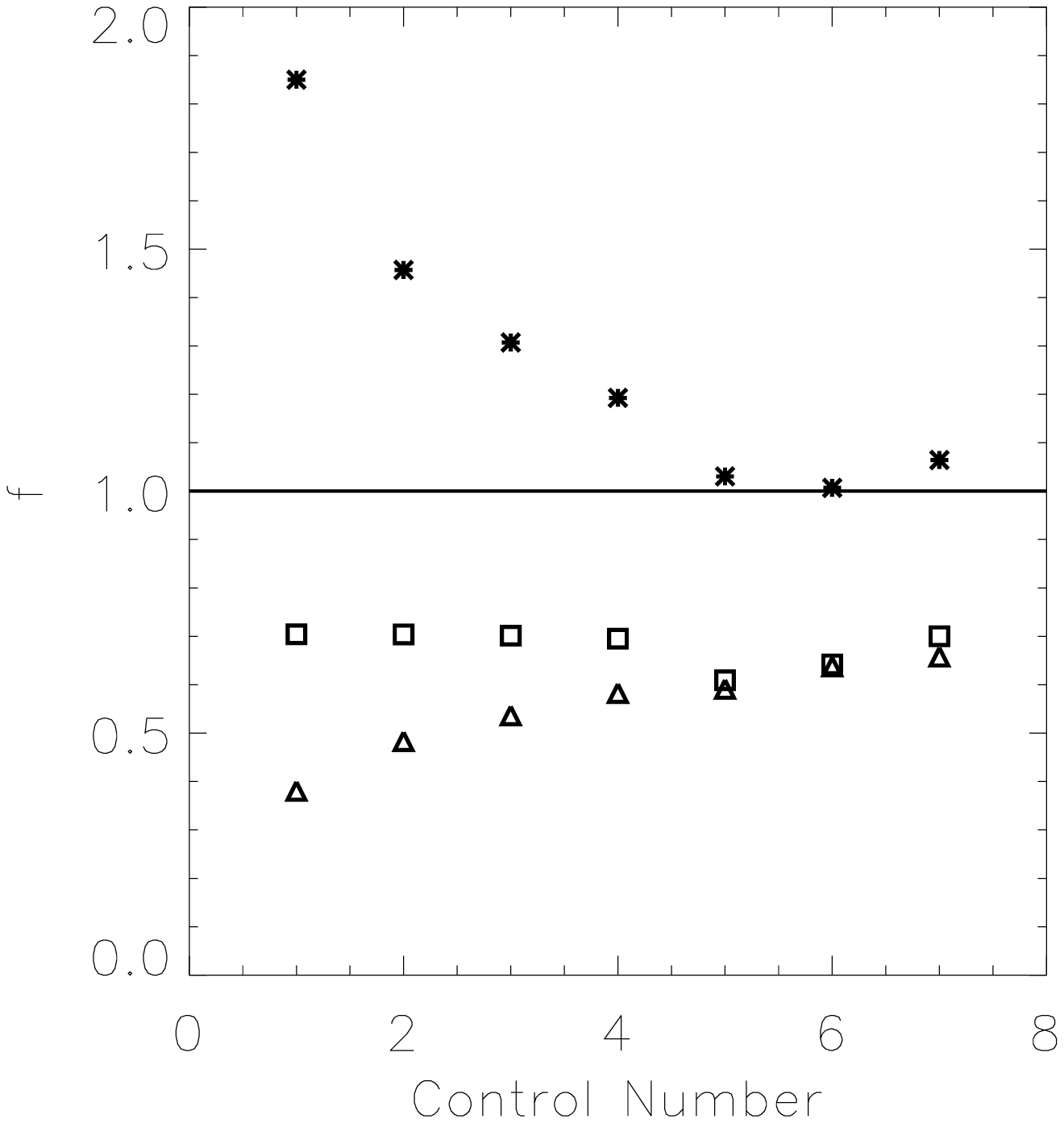}}
\hspace*{-0.2cm}
\caption{Control efficiency $C_{e}$ (asterisk) and red fractions ($u-r >2$, open triangles) 
of all analysed control samples, defined with numbers from 1 to 7 (Table~\ref{tablacontrol}). 
Red galaxy fractions computed  for each corresponding 
galaxy pair samples are also shown (open squares). 
The horizontal line corresponds to have no excess of red galaxy pairs with respect to the CS.  }
\label{fracciones-red}
\end{figure}

In order to quantify the performance of the building up process of a suitable control sample, we
estimate the  control efficiency $C_{e}$, as the ratio between the  red fraction 
of galaxies in  pairs and that of a given  control sample. 
As shown in Fig.~\ref{fracciones-red},  the efficiency of the CS  
improves from the first Control 1 to Control 5.
We can also see that Control 6 has similar $C_{e}$ than Control 5.
 It is interesting to note how the colour distributions of 
pairs and controls get closer as the different biases
are eliminated. In particular, $70\%$ of the bias is already 
cleaned up after imposing constraints on redshift, stellar mass and local
density environment to select the control sample (Control 3). 
Finally, just with the purpose of illustrating the importance of halo bias, 
we include in the figure the $C_{e}$ parameter for a new {\bf Control 7}  built imposing 
constraints only in redshift, stellar mass and halo mass. As figure shows, although
the halo mass contributes significantly to correct the total bias effect, 
the remainder constraints have to be considered in order to build a suitable control sample.



\section{An example: the mass-metallicity relation}

The mass-metallicity relation (MZR)  is a well-determined correlation 
between these two parameters which holds from ellipticals to dwarf  galaxies 
(e.g. Lequeux et al. 1979; Tremonti et al. 2004; Savianne et al. 2008). 
Recently, many authors have studied  the MZR for galaxies in pairs finding that they
tend to deviate from the mean MZR of their respective control samples 
(Kewley et al. 2006; Micheal-Dansac et al. 2008; Ellison et al. 2008).

Taking into account,  the possible biases suggested by our work in the construction of control samples 
(see also Barton et al. 2007), we analysed their impact on  the MZR of galaxies pairs in the 
Millennium Simulation. 

\begin{figure}
\resizebox{8cm}{!}{\includegraphics{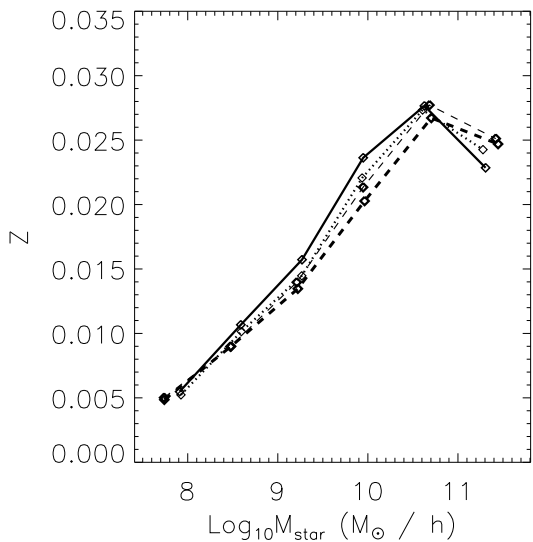}}
\hspace*{-0.2cm}
\caption{Mass-Metallicity relation for galaxy pairs (solid line),  {\bf Control 2} (dashed thick line),
 {\bf Control 3} (dashed thin line) and  {\bf Control 5} (dotted line).}
\label{MMgas}
\end{figure}

We define the metallicity parameter, Z,  as the mass in metals in 
the gas-phase component (provided by the SAM)  normalized by the cold gas mass. 
 Because this relation requires the comparison of galaxies with similar stellar components, 
we start from the estimation of the MZR for
   {\bf Control 2} (where constraints on redshift and $M_*$  have been applied).
 Nevertheless, we note that the MZR estimated from  {\bf Control 1} (which
has redshift and luminosity constraints) yields similar results.
As it can be seen from Fig.~\ref{MMgas}, galaxies in pairs (solid line) determine a 
significant different MZR compared to galaxies in {\bf Control 2} (dashed thick line), 
trend which is mainly stressed
  for stellar masses larger than $\approx 10^{9.5}  {\rm h^{-1}} M_{\odot}$.
However, an important change is obtained when  the environmental bias is corrected.
The  MZR for {\bf Control 3} (dashed thin line) 
approaches  that of  galaxy pairs as it can be appreciated from
Fig.~\ref{MMgas}. Note that this result is not modified by introducing
the constraint on morphology by using the {\bf Control 4} (not included in 
the figure for the sake of simplicity). Finally, we get the closest agreement  
between the control  and  pair MZRs when the halo mass bias
is corrected by {\bf Control 5} (dotted line).
 
For  stellar masses smaller than $\approx 10^{9.5} {\rm h^{-1}} M_{\odot}$,
 the MZRs for control samples always match closely that of  galaxy pairs.
We note that this agreement is independent of galaxy type composition  which is very different between the
two samples in this mass range.
This suggests that the Millennium treatment of galaxy types is not important.
 The clue to understand this behaviour is  given by the halo mass size. 
We find that small stellar mass systems in the control samples
live preferentially in small haloes, 
while larger stellar systems tend to  inhabit  larger haloes.
We estimated that $87\%$ of {\bf Control 2} galaxies with small stellar masses live in 
 dark matter haloes with  masses lower than $10^{12} \, {\rm h^{-1}} M_{\odot}$, while this percentage 
decreases  to $56\%$ for galaxies with larger stellar masses. A similar trend is observed 
for galaxy pairs with a $75\%$ and  $35\%$ of small and large stellar systems, respectively,
living in small dark matter haloes. 
Hence, it is only at the high stellar mass end where there is a larger 
difference in  halo composition between the control and pair samples (see 
Fig.~\ref{bias1}).

\begin{figure}
\resizebox{8cm}{!}{\includegraphics{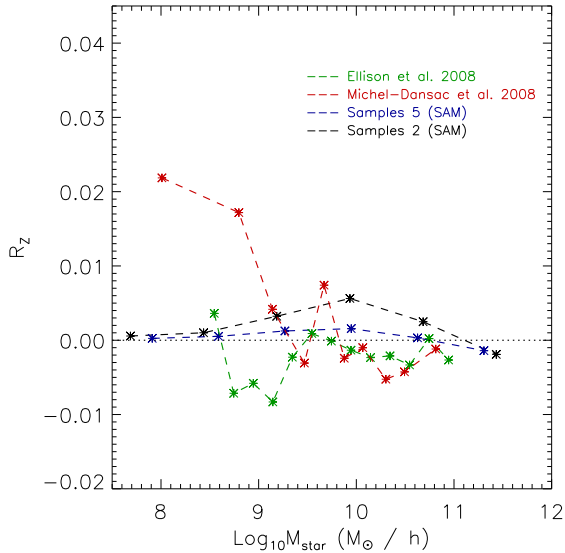}}
\hspace*{-0.2cm}
\caption{The pair metallicity excess, $R_{Z}$, defined as the difference between pair and CS metallicities normalized 
by the pair value, as a function of stellar masses. Red and green lines represent the SDSS-DR4 
values computed with $O/H$ abundances kindly provided by Michel-Dansac et al. (2008) and Ellison et al. (2008), respectively. 
 Analogously, mock  $R_{Z}$ values for Samples 2 and  5 are represented 
by  black and blue lines, respectively.}
\label{res}
\end{figure}

These findings suggest that observational results on the MZR for galaxy pairs 
might be affected by  biases principally
at the high stellar mass end. In order to test this hypothesis, we introduce the 
  pair metallicity excess parameter,  $R_{Z}$, defined as 
 the difference of pair and CS metallicities normalized by the pair value, and calculate it 
for observational and theoretical data. 
Fig.~\ref{res} shows the observational $R_{Z}$, computed with the $O/H$  abundances of SDSS-DR4 galaxies 
generously provided by Micheal-Dansac et al. (2008) and Ellison et al. (2008) (hereafter MD08 and E08, respectively),  
as a function of the stellar mass. 
It is interesting to note that even when they made a different selection of their pair and control samples,  
both trends in Fig.~\ref{res} (red and green lines) are  appreciably consistent. They 
find  that for intermediate and large stellar mass systems, the metallicities of galaxy pairs slightly tend to have 
 lower values  respect to those in their respective isolated galaxy samples. 
 This trend reverses (at least in the case of MD08) for smaller stellar masses,
with galaxy pairs showing an excess of metals respect to their isolated counterparts.

 Fig.~\ref{res} also shows the pair metallicity excess of mock galaxies computed with 
our pair and control galaxies of Samples 2 and 5 (black and blue lines, respectively). 
{\bf We warn that when comparing  semi-analytical and observational   
$R_{Z}$ values some issues must be born in mind}. First, while the $O/H$  abundances can be estimated 
for SDSS galaxies, only a mean gas-phase metallicities can be obtained from the SAM. Second, 
because of the reduced size of the spectroscopic fiber, metallicities of SDSS galaxies tend to be nuclear 
(depending on the galaxy size), however, SAM provides a mean value of the global metallicity of galaxies. 
Finally, and probably the main reason, SAM does not include the physics of interactions, consequently, mock ZMRs 
 certainly cannot reflect the tidal trace as in observations. 
{\bf Nevertheless, these reasons do not invalidate the comparison since we are always 
evaluating the excess with respect to the appropriate control sample which shares the same limitations.}  
The  inspection of $R_{Z}$ for mock galaxies  shows that our Samples 2 
(black line of Fig.~\ref{res}) exhibit a different metal content in galaxy with and without a near
companion, with higher metallicities in galaxy pairs at intermediate stellar masses. 
However, after correcting morphology, local density environment and halo mass biases 
as done for Samples 5 (blue line), these differences are significantly removed. 
{\bf The comparison of  $R_{Z}$ for Samples 2 and 5 shows that a biased selection might affect the interpretation of ZMRs  
only at intermediate stellar masses, suggesting that the observed values could be even lower than
reported. Our results support the trends detected by MD08 and E08  
at low and high stellar mass ends, where the theoretical $R_{Z}$ values are almost negligibly.}

\section{Conclusions.}

In this work, we  analyse how to build up a suitable control sample in order to isolate the effects 
of interactions on the colour  and star formation activity distributions of galaxies in pairs. 
We took profit on the fact that the SAMs do not include the effects of 
interactions, so that, mock galaxies with and without a close companion should have similar
colour distribution and star formation activity if the only difference between them is the presence of a companion.

 We found that a control sample selected by imposing their members to have only the same
luminosity and redshift distributions than galaxies in a pair sample ends up formed
by a galaxy population that differs in gas content, stellar masses, morphology, environment and
dark matter haloes. Because of these biases,   
galaxies in pairs seem to be artificially older, gas-poorer,  bulge-dominated and tend 
to inhabit higher local density regions and higher DM haloes when comparing
with their isolated counterparts in this basic control sample. The galaxy pair MZR is
 also affected by these bias selection.

Hence, if a control sample is not cleaned from these biases, then the confrontation with
galaxy pairs could yield spurious results. We systematically took each of these biases 
into account to correct the control samples and finally
get one with  the same colour distribution and star formation activity as galaxies in pairs. 
This control sample also has similar gas fraction and 
mean stellar-mass weighted ages to those of galaxies in pairs.

We found that the differences between the control and pairs samples diminished by $70\%$ by  
considering constraints on  redshift, stellar mass and local  density.   
We also showed that the effects of dark matter haloes could be  overestimated in the SAM 
so that our estimations should be considered upper limits. 

 Some of the constraints we have used, such as galaxy types or halo mass, are difficult to estimate 
observationally. However, via the theoretical analysis  of their effects we 
could assess how relevant they are for the study of pair galaxies. We conclude
that, on one hand, galaxy type bias is the less important one compared
to the environment and mass ones. On the other hand,  halo mass bias could
be very significant as previously reported (Barton et al 2007) but by
taking into account  environment bias, its effects are importantly mitigated. 
Our comprehensive study of mock galaxies  showed that 
 a suitable control sample for isolating the effects of interactions should
be built by imposing constraints on redshift, stellar mass, local environment, morphology and halo mass.
Only when these criteria are applied, the differences found in the bimodal colour distribution
(MZR and star formation activity) could be directly associated to the effects of interactions.

\section*{Acknowledgments}
We thank Simon White for his useful comments and suggestions, and 
  Sara Ellison and Leo Michel-Dansac for providing  electronic access to their data. 
J.P. particularly thanks Gerard Lemson for his selfless help to manage
 the Millennium Data Base. We also thank the referee for thoughtful comments that
helped to  improve this paper. This work was partially supported by the
Consejo Nacional de Investigaciones Cient\'{\i}ficas y T\'ecnicas, AGENCIA 
(Pict 32342 (2005) and Pict 245 (2006) 
and by the European Union's ALFA-II programme, 
through LENAC, the Latin American European Network for
Astrophysics and Cosmology.

\end{document}